\documentclass[a4paper,11pt]{article}
\usepackage{pos}
\usepackage{amsmath,bm,longtable,mathrsfs,slashed}

\title{Casimir effect for fermions on the lattice}
\ShortTitle{Casimir effect for fermions on the lattice}

\manuallySeparateAuthors
\author*[a]{Katsumasa Nakayama}
\author[]{, and}
\author[b]{ Kei Suzuki}

\affiliation[a]{RIKEN Center for Computational Science, Kobe 650-0047, Japan}
\affiliation[b]{Advanced Science Research Center, Japan Atomic Energy Agency (JAEA), Tokai 319-1195, Japan}

\emailAdd{katsumasa.nakayama@riken.jp}
\emailAdd{k.suzuki.2010@th.phys.titech.ac.jp}

\abstract{
The conventional Casimir effect has been studied in the continuous spacetime, but to elucidate its counterpart in the lattice space is an important subject. 
Here, we discuss various types of Casimir effects for quantum fields on the lattice.
By using a definition of the Casimir energy on the lattice, we show that the Casimir effect for the Wilson fermion is similar to that for the continuous Dirac fermion.
We apply our definition to an effective Hamiltonian describing Dirac semimetals, such as Cd$_{3}$As$_{2}$ and Na$_{3}$Bi, and find an oscillatory behavior of the Casimir energy as a function of film thickness of semimetals.
We also study contributions from Landau levels under magnetic fields and the Casimir effect for nonrelativistic particle fields on the lattice.
}

\FullConference{%
 The 39th International Symposium on Lattice Field Theory, LATTICE2022
  8th-13th August, 2022,
  Bonn, Germany.
}

\begin{document}
\maketitle

\section{Introduction}\label{sec-1}
The Casimir effect~\cite{Casimir:1948dh} was originally predicted as a quantum phenomenon induced from the zero-point energy of photon fields under the existence of spatial boundary conditions, which is nothing but a finite-volume effect for the zero-point energy.
As a result, an attractive force between two parallel metal plates, the Casimir force, has been experimentally established~\cite{Lamoreaux:1996wh,Bressi:2002fr}.
Since the origin of the Casimir effect is the zero-point energy of a quantum field, various types of particle fields, including scalar/fermion/gauge fields, should exhibit analogous Casimir effects.

\begin{figure}[b!]
    \centering
    \begin{minipage}[t]{0.5\columnwidth}
    \includegraphics[clip,width=1.0\columnwidth]{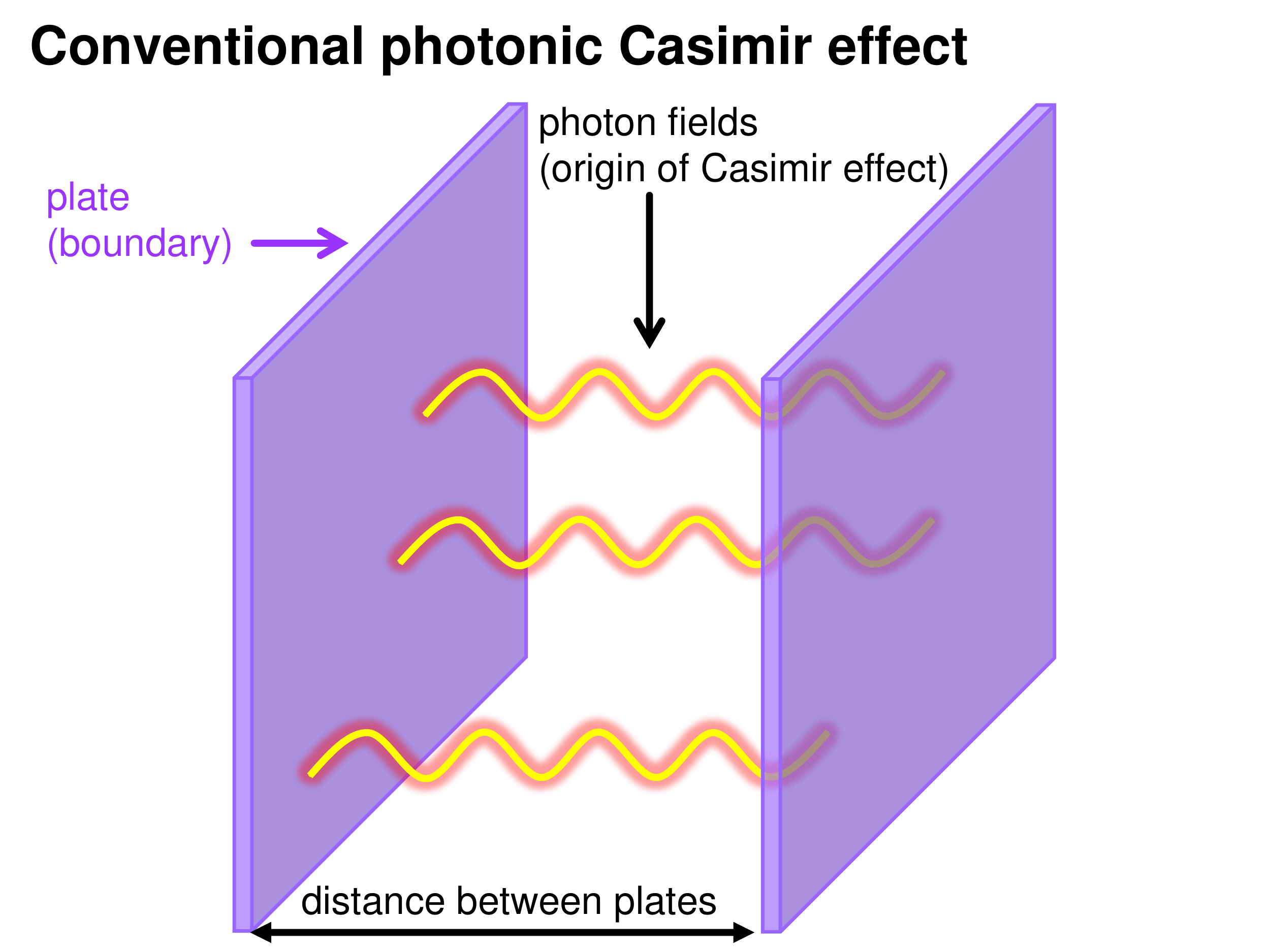}
    \end{minipage}%
    \begin{minipage}[t]{0.5\columnwidth}
    \includegraphics[clip,width=1.0\columnwidth]{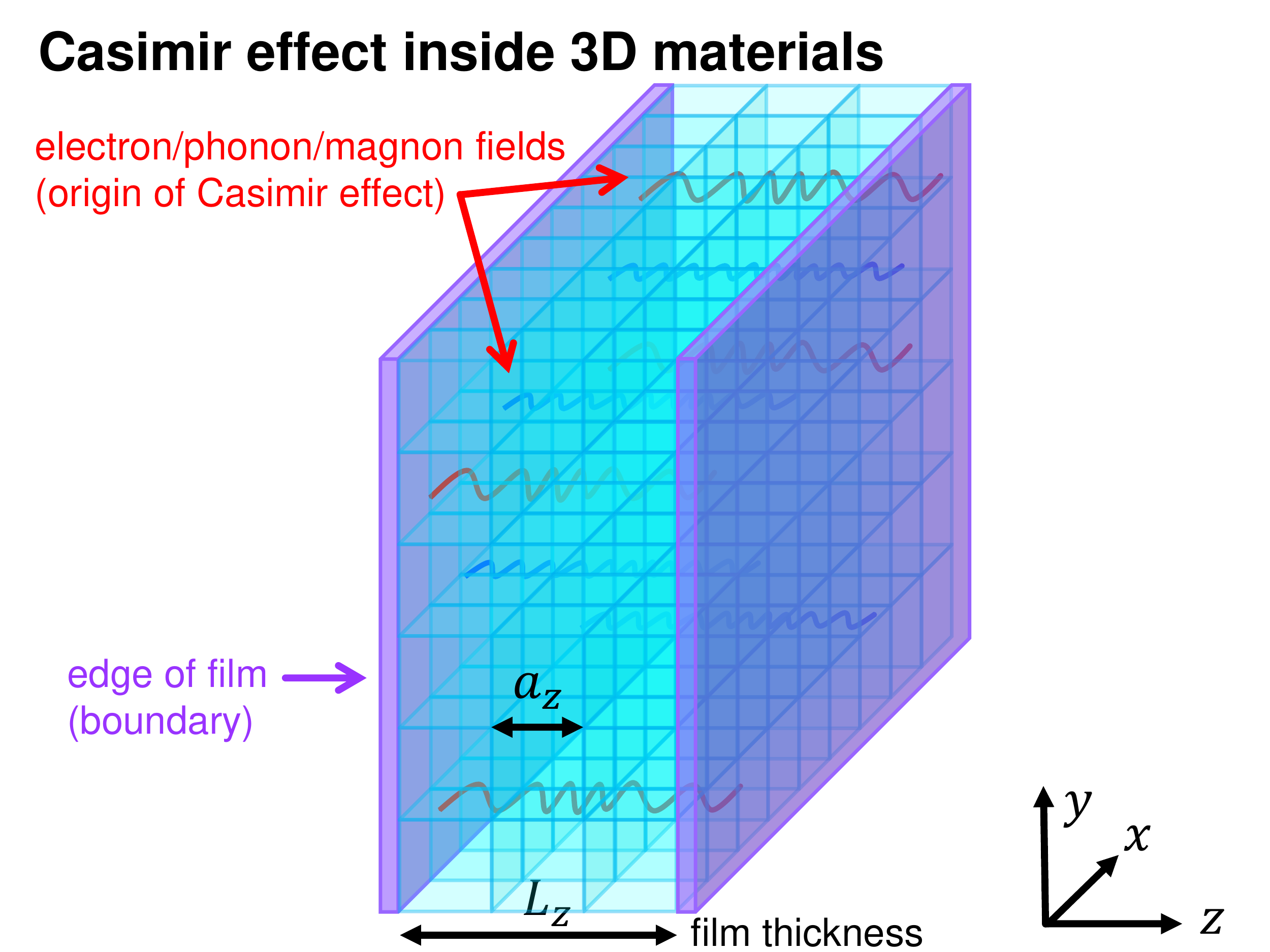}
    \end{minipage}
  \caption{Left: Conventional Casimir effect for photon fields.
Right: Casimir effects for quantum fields on the lattice.}
    \label{fig:comparison}
\end{figure}

In order to calculate the Casimir effect in the continuous spacetime, one needs to remove the ultraviolet divergence of the zero-point energy, by using mathematical techniques.
On the lattice spacetime, such a divergence does not exist, so that physical quantities resulting from the Casimir effect are theoretically well-defined.  
Quantum fields on the lattice space are common degrees of freedom in solid state physics, such as electrons, phonons, and magnons, and these fields can also induce the Casimir effects under boundary conditions.
In Fig.~\ref{fig:comparison}, we compare the conventional photonic Casimir effect (the left) and the Casimir effects realized inside thin films of three-dimensional (3D) materials (the right).
Thus, the Casimir effect inside materials {\it intrinsically} exists as long as a system is ``finite-size," and particularly the geometries with boundaries in the one direction, such as thin films of 3D materials, narrow nanoribbons of 2D materials, and short nanochains of 1D materials, are analogous to the conventional setup.
These Casimir effects automatically contribute to the internal pressure and other thermodynamic quantities of the system.

The Casimir effect on the lattice is calculated using a lattice regularization scheme~\cite{Actor:1999nb,Pawellek:2013sda,Ishikawa:2020ezm,Ishikawa:2020icy,Nakayama:2022ild,Nakata:2022pen,Mandlecha:2022cll,Nakayama:2022fvh,Swingle:2022vie}.
The Casimir effect on the lattice was investigated for relativistic scalar fields~\cite{Actor:1999nb,Pawellek:2013sda}, relativistic fermion fields~\cite{Ishikawa:2020ezm,Ishikawa:2020icy,Mandlecha:2022cll,Swingle:2022vie}, and nonrelativistic particle fields~\cite{Nakayama:2022ild} (specifically, magnon fields in antiferromagnets and ferrimagnets~\cite{Nakata:2022pen} and electron fields in Dirac/Weyl semimetals~\cite{Nakayama:2022fvh}).
For example, we consider the three dimensional lattice with a length $L_z=a_zN_z$ between boundary conditions imposed in the $z$ direction, where $a_i$ and $N_i$ ($i=x,y,z$) are the lattice spacing and the number of lattice links, respectively (see the right of Fig.~\ref{fig:comparison}). 
Then, the Casimir energy per the surface area $L_xL_y=a_xa_yN_xN_y$ (which precisely should be called the Casimir ``surface density") is defined with the lattice regularization as follows:
\begin{subequations}
\begin{align}
E_\mathrm{Cas} &\equiv E_0^\mathrm{sum}(N_z) - E_0^\mathrm{int}(N_z), \label{eq:def_cas} \\
E_0^\mathrm{sum}(N_z) &= \sum_j \int_\mathrm{BZ} \frac{d^2(a_ik_\perp)}{(2\pi)^2} \left[ \pm \frac{1}{2}  \sum_{n} |\omega_{k_\perp,n,j}|  \right],  \label{eq:def_E0sum}\\
E_0^\mathrm{int}(N_z) & =  \sum_j \int_\mathrm{BZ} \frac{d^2(a_ik_\perp)}{(2\pi)^2} \left[\pm \frac{N_z}{2} \int_\mathrm{BZ} \frac{d(a_zk_z)}{2\pi} |\omega_{{\bf k},j}| \right]. \label{eq:def_E0int}
\end{align}
\end{subequations}
Thus, the Casimir energy $E_\mathrm{Cas}$ is defined as the difference between the zero-point energies $E_0^\mathrm{sum}$ (with discrete eigenvalues $\omega_{k_\perp,n,j}$ labeled by $n$) and $E_0^\mathrm{int}$ (with continuous eigenvalues $\omega_{{\bf k},j}$).
The momentum integrals with respect to the continuous spatial momenta ${\bf k} =(k_x,k_y,k_z)$ are taken within the first Brillouin zone (BZ), where $d^2(a_ik_\perp) \equiv d(a_xk_x)d(a_yk_y)$.
The discrete-momentum sum $\sum_n$ is also taken within the first BZ.
$\sum_j$ is the sum of possible eigenvalues labeled by $j$ (which physically corresponds to particle/antiparticle and spin degrees of freedom).
The sign of $\pm$ is fixed as $+$ for boson fields and $-$ for fermion fields.
The factor of $1/2$ is a feature of the zero-point energy.
The absolute value is needed to correctly take into account negative-energy eigenvalues of relativistic fields.
As some examples of boundary conditions, $k_z$ in $E_0^\mathrm{sum}$ is discretized as
\begin{subequations}
\begin{align}
&a_zk_z \to \frac{2n\pi}{N_z} && (n=0,\cdots, N_z-1 \ \mathrm{or} \ 1,\cdots, N_z) && (\mathrm{periodic}), \label{eq:def_PBC}\\
&a_zk_z \to \frac{(2n+1)\pi}{N_z} && (n=0,\cdots, N_z-1 \ \mathrm{or} \ 1,\cdots, N_z) && (\mathrm{antiperiodic}), \label{eq:def_APBC}\\
&a_zk_z \to \frac{n\pi}{N_z} && (n=0,\cdots, 2N_z-1 \ \mathrm{or} \ 1,\cdots, 2N_z) && (\mathrm{phenomenological}). \label{eq:def_phenoBC}
\end{align}
\end{subequations}
For the phenomenological boundary~(\ref{eq:def_phenoBC}), we have to replace the sum in Eq.~(\ref{eq:def_E0sum}) as $\sum_{n} \to \frac{1}{2} \sum_n$.
Note that the two types of ranges of $n$, the ranges starting from $n=0$ and that from $n=1$, are equivalent to each other.

The definition (\ref{eq:def_cas}) is similar to that in the continuous spacetime (as defined by Casimir~\cite{Casimir:1948dh}), but on the lattice there is no ultraviolet divergence in each term by the existence of the lattice cutoff.
Therefore, we can obtain the solution from the exact summation and the numerical integration (or analytic integration, if possible).
In these proceedings, we show various types of Casimir effects on the lattice, by substituting some types of $\omega_{{\bf k},j}$ into the definition (\ref{eq:def_cas}).

\section{Wilson fermions}\label{sec-2}
First, as an instructive example, we demonstrate the Casimir effect for the Wilson fermions (see Refs.~\cite{Ishikawa:2020ezm,Ishikawa:2020icy,Mandlecha:2022cll} for details).
Since the low-energy (infrared) behavior of the conventional Wilson fermion simulate that of the Dirac fermions, we expect a similar Casimir effect if the Casimir energy is determined by its infrared dynamics of quantum fields.

The dispersion relations of 3D Wilson fermions~\cite{Wilson:1975,Wilson:1977} are given as
\begin{equation}
\omega_\pm^\mathrm{Wilson} = \pm \frac{\hbar c}{a_i} \sqrt{\sum_i^{x,y,z} \sin^2 a_ik_i + \left[ r \sum_i^{x,y,z}  (1 - \cos a_ik_i) + a_im_f \right]^2}.
\end{equation}
For simplicity, we set the natural unit ($\hbar=1$, $c=1$), $a \equiv a_x=a_y=a_z$, and $r=1$.

\begin{figure}[tb!]
    \centering
    \begin{minipage}[t]{0.5\columnwidth}
    \includegraphics[clip,width=1.0\columnwidth]{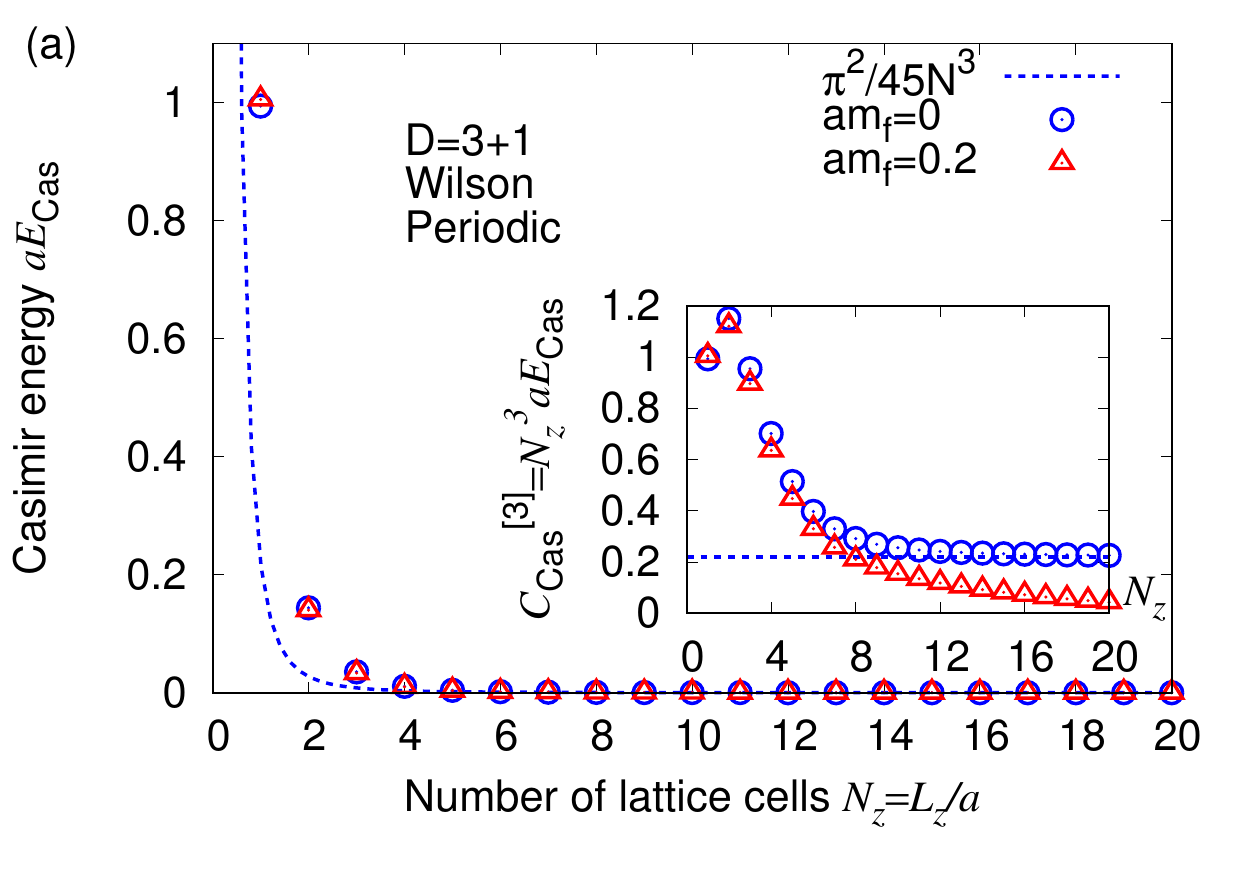}
    \end{minipage}%
    \begin{minipage}[t]{0.5\columnwidth}
    \includegraphics[clip,width=1.0\columnwidth]{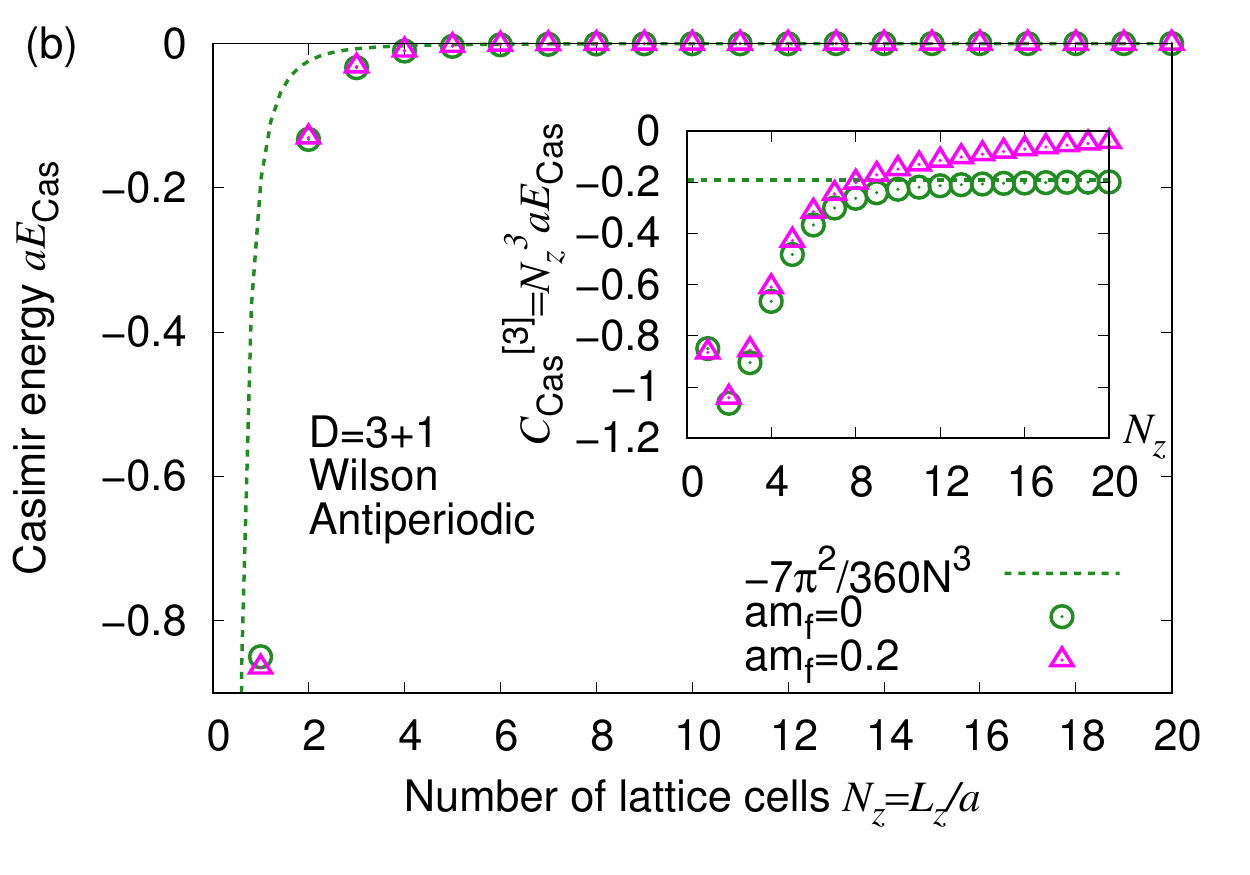}
    \end{minipage}
    \caption{Casimir energy and Casimir coefficients for Wilson fermions in the 3+1 dimensions (see Refs.~\cite{Ishikawa:2020ezm,Ishikawa:2020icy} for details).
(a) Periodic boundary condition
(b) Antiperiodic boundary condition.
}
    \label{Fig:Wilson}
\end{figure}

Figures~\ref{Fig:Wilson}(a) and (b) show the dimensionless Casimir energy $aE_\mathrm{Cas}$ and the dimensionless Casimir coefficient defined as $C_\mathrm{Cas}^{[d]} \equiv N_z^d a E_\mathrm{Cas}$ with the periodic and antiperiodic boundary conditions.
From these figures, the Casimir effect for massless Wilson fermions ($am_f=0$) is consistent with the analytic solutions for the Dirac fermion (dotted lines) in the larger $N_z$, while the difference in the smaller $N_z$ reflects the ultraviolet lattice effect of Wilson fermions.
When we switch on a fermion mass ($am_f=0.2$), we find that the Casimir energy in the larger $N_z$ is suppressed, which is similar to the property well known in the Casimir effect for massive particle fields.

\section{Effective Weyl electrons in Weyl semimetals}\label{sec-3}
\begin{figure}[tb!]
    \centering
   \begin{minipage}[t]{0.33\columnwidth}
    \includegraphics[clip,width=1.0\columnwidth]{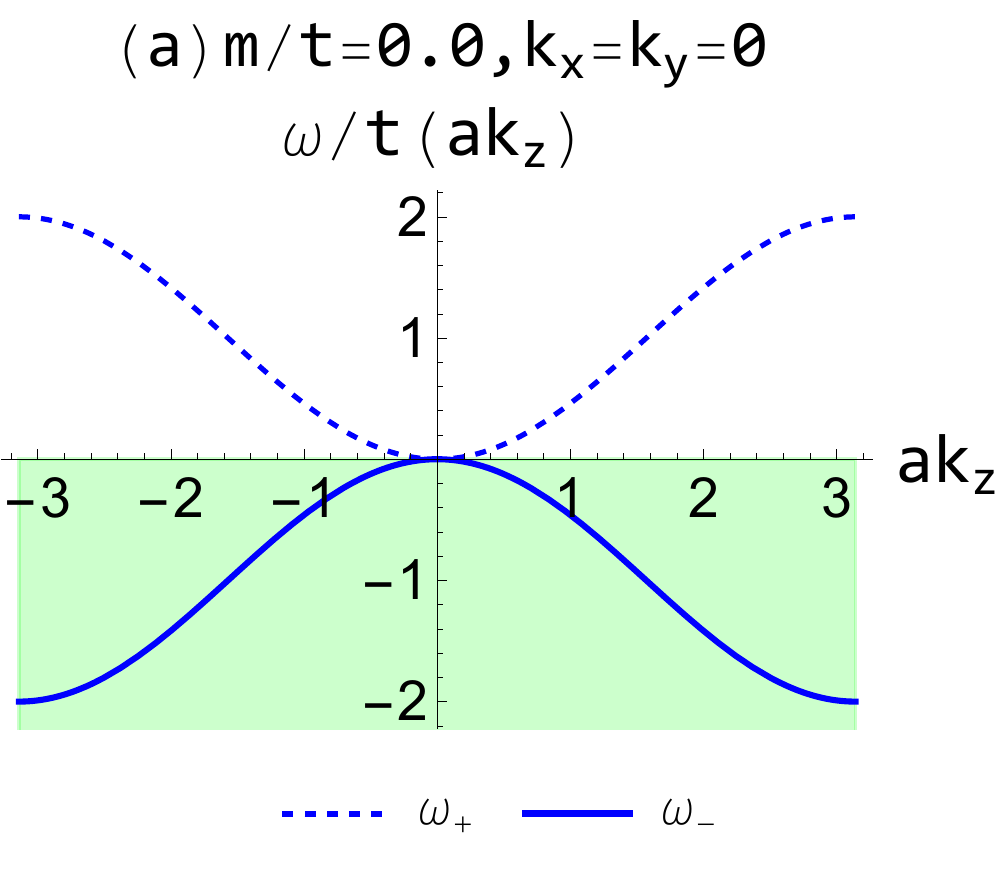}
    \end{minipage}%
    \begin{minipage}[t]{0.33\columnwidth}
    \includegraphics[clip,width=1.0\columnwidth]{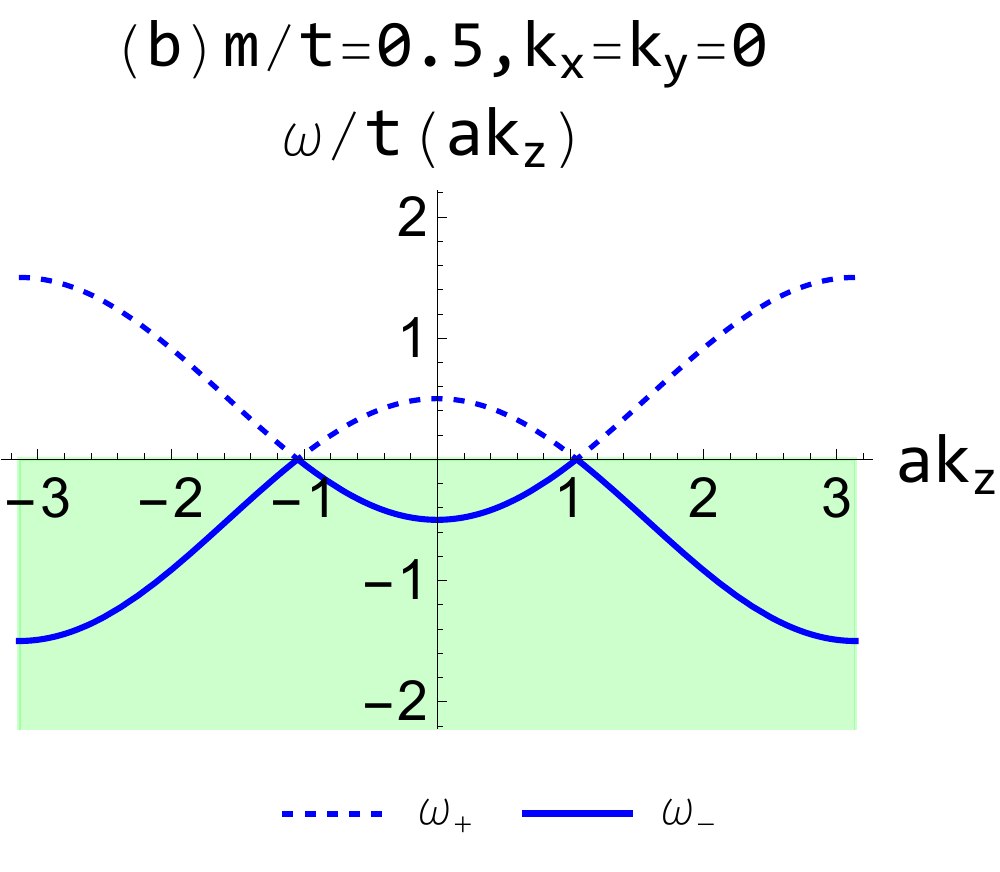}
    \end{minipage}%
    \begin{minipage}[t]{0.33\columnwidth}
    \includegraphics[clip,width=1.0\columnwidth]{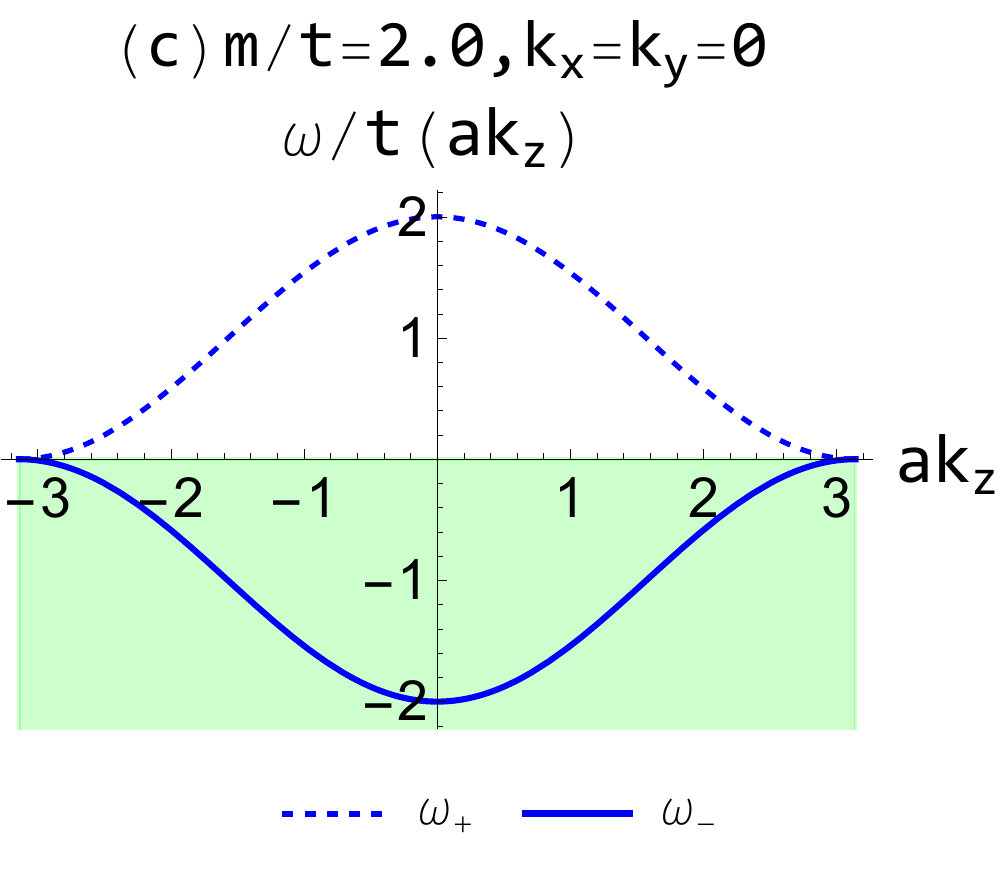}
    \end{minipage}
    \begin{minipage}[t]{0.5\columnwidth}
    \includegraphics[clip,width=1.0\columnwidth]{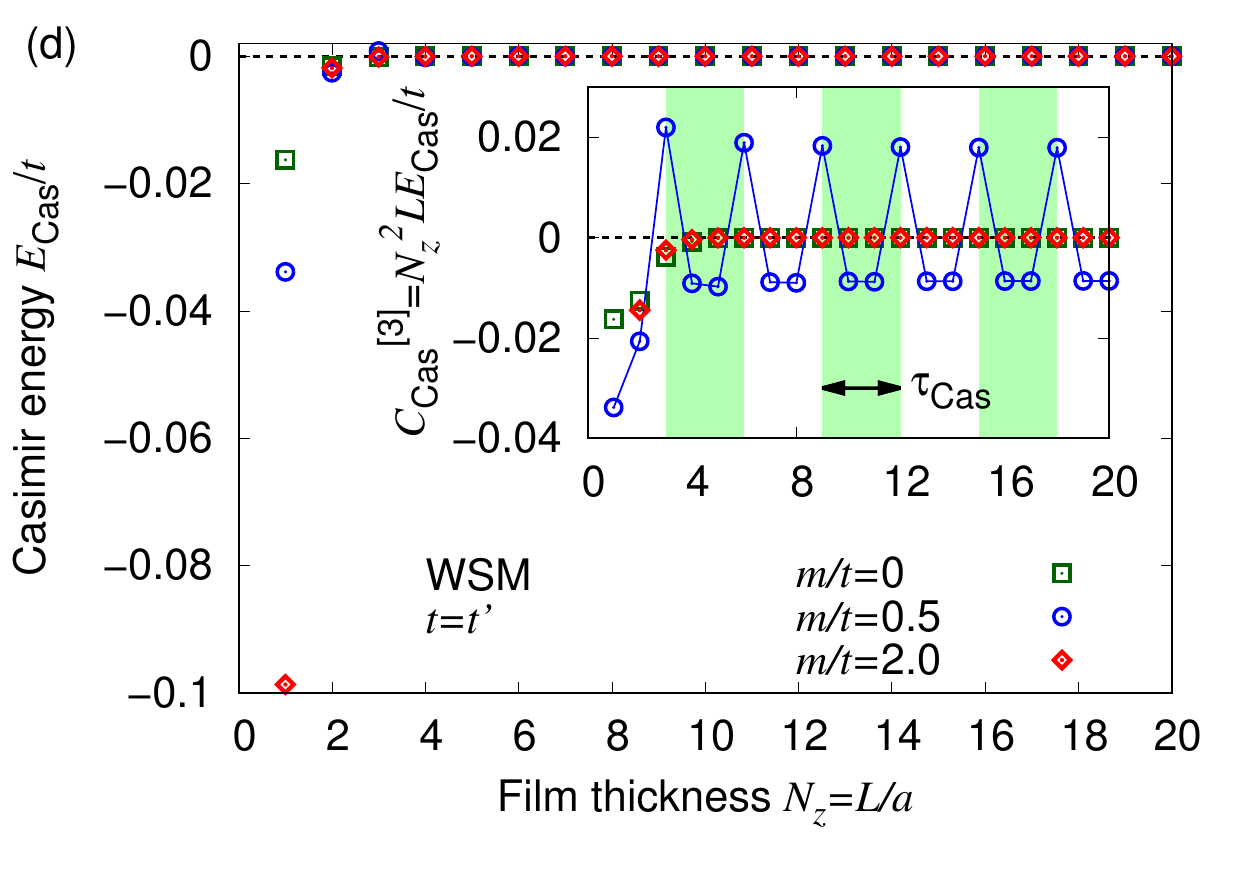}
    \end{minipage}
    \caption{(a)-(c) Dispersion relations of Weyl fermions in Weyl semimetals described by Eq.~(\ref{eq:WSMtoy}), where the model parameter is fixed as $m/t=0.0$, $0.5$, or $2.0$.
(d) Casimir energy for Weyl fermions with the phenomenological boundary (see Ref.~\cite{Nakayama:2022fvh} for details).
}
    \label{toy}
\end{figure}

As an example of interesting behaviors of the Casimir effect on the lattice, we here demonstrate an oscillatory behavior of the Casimir energy.
We consider the dispersion relations obtained from an effective Hamiltonian of time-reversal-symmetry breaking 3D Weyl semimetals, given as~\cite{Yang:2011}
\begin{align}
\omega_\pm^\mathrm{WSM} = \pm \sqrt{ t^2 \sum_i^{x,y} \sin^2 a_ik_i + \left[m - t^\prime \sum_i^{x,y,z} (1-\cos a_ik_i) \right]^2}. \label{eq:WSMtoy}
\end{align}
This band structure is characterized by the three parameters, $t,  t^\prime$, and $m$.
For simplicity, we assume $t=t^\prime$.
In the parameter region of $0<m/t<2$, these dispersion relations are distributed across the zero (Fermi) energy, and then the crossing points are regarded as the two Weyl points at $(k_x,k_y,k_z) = (0,0,\pm k_\mathrm{WP})$ with $k_\mathrm{WP} = \frac{1}{a_z} \arccos(1-m/t)$ [e.g., see Fig.~\ref{toy}(b)].
At $m/t=0$ and $2$, the eigenvalues just touch the Fermi level without forming Weyl points [see Figs.~\ref{toy}(a) and (c)].
In the regions of $m/t<0$ and $m/t>2$, the eigenvalues are separated from the Fermi level.

Figure~\ref{toy}(d) shows the dimensionless Casimir energy $E_\mathrm{Cas}/t$ at $m/t=0.0$, $0.5$, or $2.0$ and the dimensionless Casimir coefficient defined as $C_\mathrm{Cas}^{[3]} \equiv N_z^3E_\mathrm{Cas}/t$ with the phenomenological boundary~(\ref{eq:def_phenoBC}).
We find that, at $m/t=0.5$, the Casimir energy oscillates with a period $\tau_\mathrm{Cas}$.
This period is characterized only by the distance between the two Weyl points (WPs)~\cite{Nakayama:2022fvh}:
\begin{align}
\tau_\mathrm{Cas} = \frac{\pi}{a_z k_\mathrm{WP}} . \label{eq:period}
\end{align}
For $m/t=0.5$, the Weyl points are located at $a_z k_z = \pm a_z k_\mathrm{WP} = \pm \pi/3$, so that the corresponding period is $\tau_\mathrm{Cas} = 3$.
At $m/t=0.0$ and $2.0$, since there are no Weyl points, the Casimir energy does not oscillate.
Thus, the oscillation of Casimir energy is useful as a signal of Weyl points (or nodes) at finite momentum.

\section{Effective Dirac electrons in Dirac semimetals}\label{sec-4}
The concepts shown in the Sec.~\ref{sec-3} can be applied to realistic Dirac/Weyl semimetals.
From an effective Hamiltonian describing 3D Dirac semimetals such as Cd$_3$As$_2$ and Na$_3$Bi, we obtain the following dispersion relations~\cite{Wang:2012,Wang:2013}:
\begin{align}
\omega_\pm^\mathrm{DSM} = \epsilon_0 \pm \sqrt{M^2+A^2(k_x^2+k_y^2)}, \label{eq:DSM}
\end{align}
where $\epsilon_0 = C_0 + C_1k_z^2 + C_2(k_x^2+k_y^2)$ and $M = M_0 +M_1k_z ^2 +M_2 (k_x^2+k_y^2)$.
For Cd$_3$As$_2$, the model parameters are given as
$A = 0.889$ eV\AA,
$C_0 = -0.0145$ eV, $C_1 = 10.59$ eV\AA$^2$, $C_2 = 11.5$ eV\AA$^2$,
$M_0 = -0.0205$ eV, $M_1 = 18.77$ eV\AA$^2$, $M_2 = 13.5$ eV\AA$^2$~\cite{Cano:2016eie}, $a_{x} = a_{y} = 12.67$ \AA$\,$, and $a_z = 25.48$~\AA~\cite{Steigmann:1968}.
For Na$_3$Bi, $A = 2.4598$ eV\AA,
$C_0 = -0.06382$ eV, $C_1 = 8.7536$ eV\AA$^2$, $C_2 = -8.4008$ eV\AA$^2$,
$M_0 = -0.08686$ eV, $M_1 = 10.6424$ eV\AA$^2$, $M_2 = 10.361$ eV\AA$^2$,
$a_{x} = a_{y} = 5.448$ \AA, and $a_z = 9.655$ \AA~\cite{Wang:2012}.
To construct the dispersion relations on the lattice, we replace $k_i^2 \to \frac{1}{a_i^2} \sin^2 ak_i$ for the term with $A$ and $k_i^2 \to \frac{1}{a_i^2} (2-2\cos ak_i)$ for the other terms.

\begin{figure}[t!]
    \centering
    \begin{minipage}[t]{0.33\columnwidth}
    \includegraphics[clip,width=1.0\columnwidth]{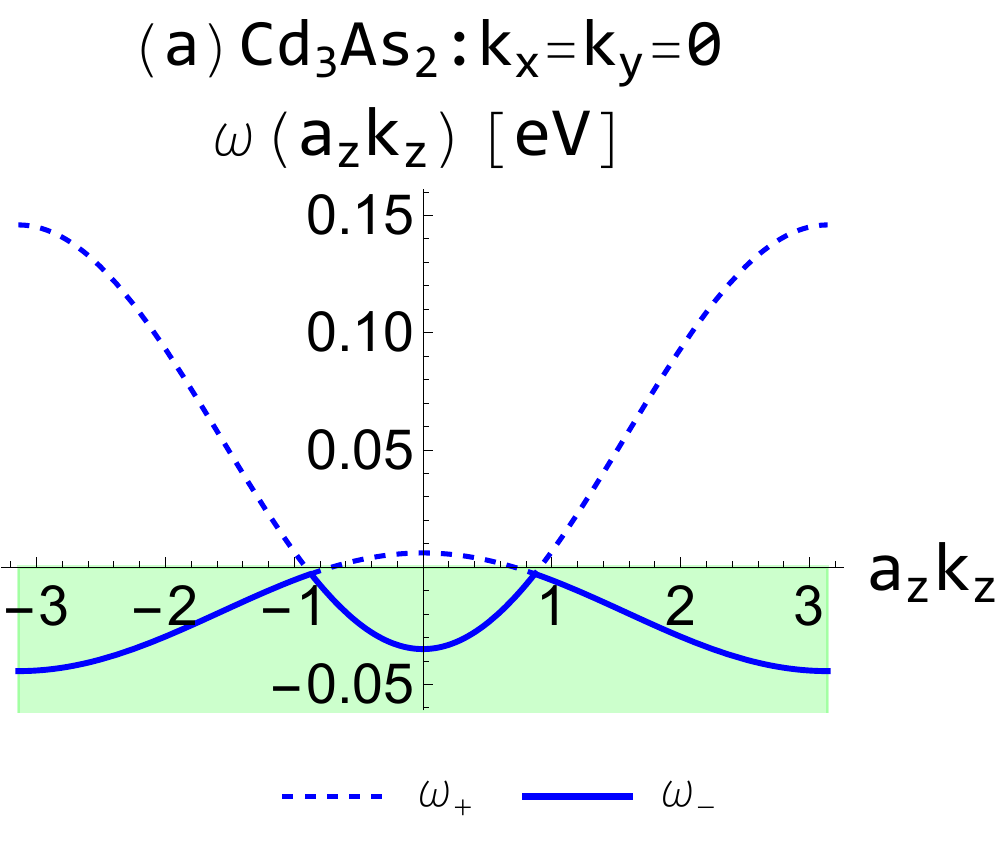}
    \end{minipage}%
    \begin{minipage}[t]{0.33\columnwidth}
    \includegraphics[clip,width=1.0\columnwidth]{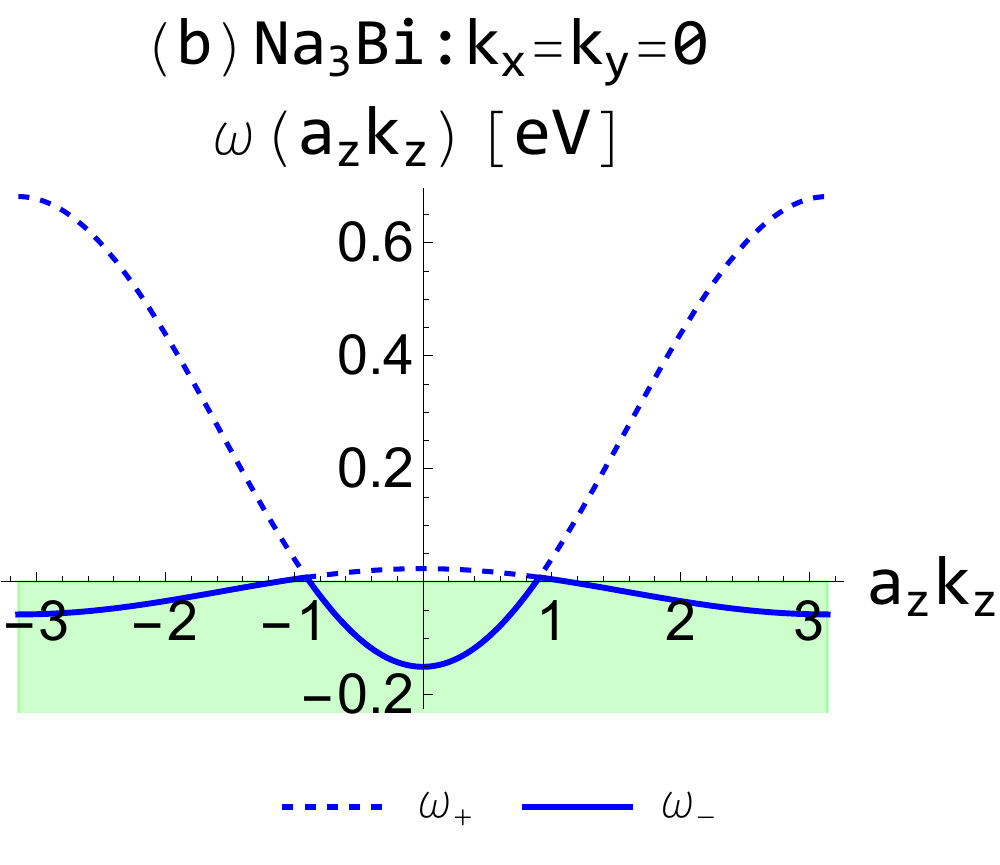}
    \end{minipage}
    \begin{minipage}[t]{0.5\columnwidth}
    \includegraphics[clip,width=1.0\columnwidth]{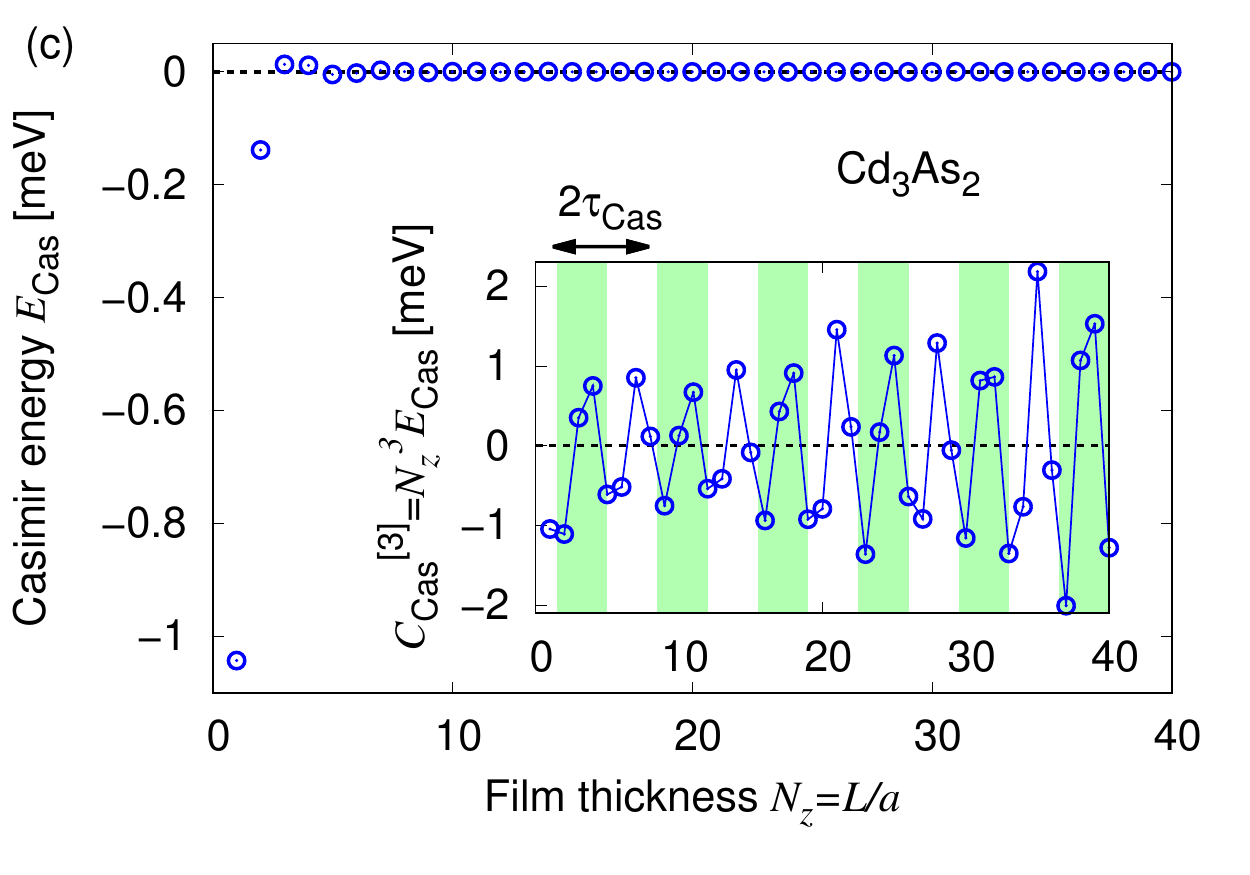}
    \end{minipage}%
    \begin{minipage}[t]{0.5\columnwidth}
    \includegraphics[clip,width=1.0\columnwidth]{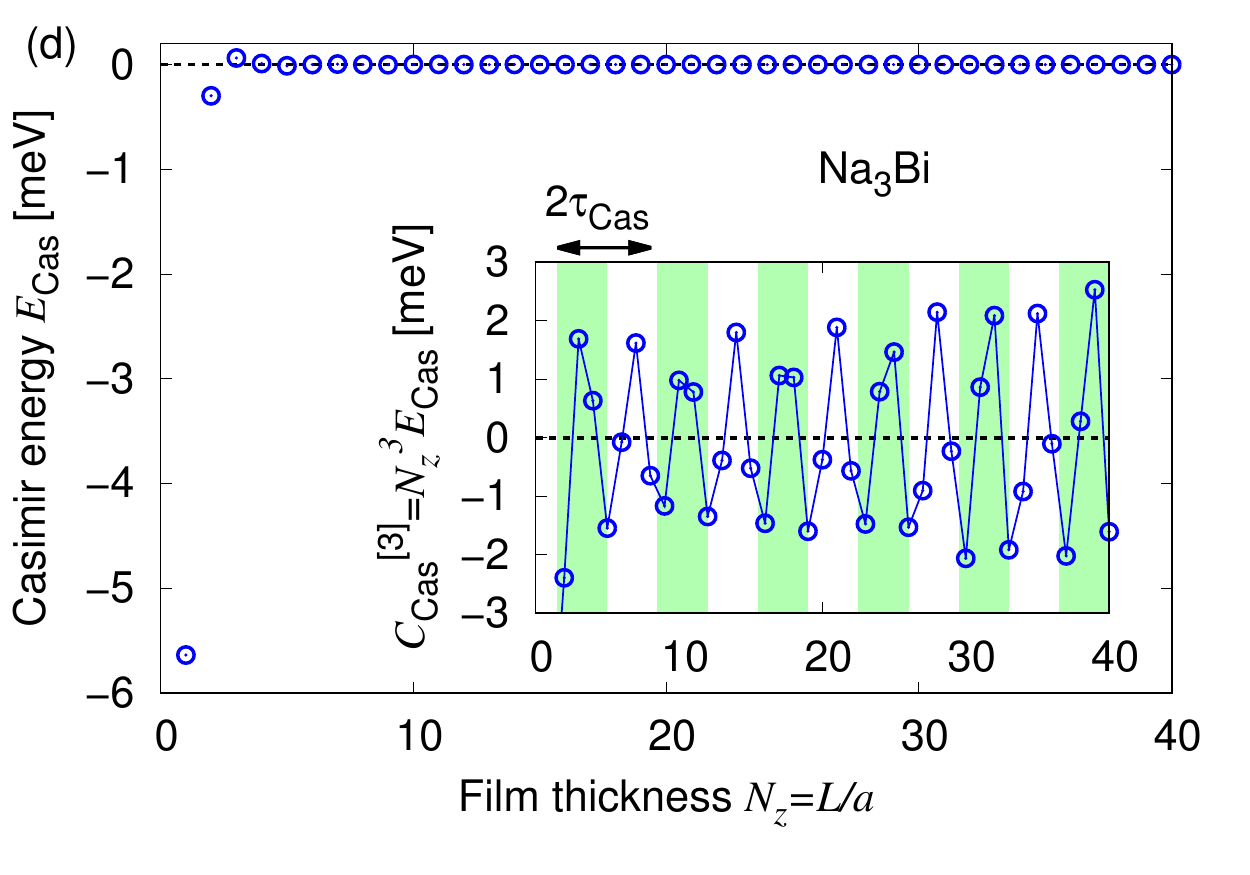}
    \end{minipage}
    \caption{(a)-(b) Dispersion relations of Dirac electrons in Cd$_3$As$_2$ or Na$_3$Bi thin films described by Eq.~(\ref{eq:DSM}).
(c)-(d) Casimir energy for Dirac electrons with the phenomenological boundary (see Ref.~\cite{Nakayama:2022fvh} for details).}
    \label{fig:DSM}
\end{figure}

Figure~\ref{fig:DSM} shows the numerical results.
For both Cd$_3$As$_2$ and Na$_3$Bi, we can find the oscillation of the Casimir energy and coefficient with a period characterized by the Dirac points (DPs).
From these results, we expect the periods $\tau_\mathrm{Cas} = \frac{\pi}{a_z k_\mathrm{DP}} \sim 3.73$ and $3.60$, which corresponds to the film thicknesses of $L_z = a_z\tau_\mathrm{Cas} \sim 9.5$ nm and $3.5$ nm, respectively.

\section{Landau levels in Weyl/Dirac semimetals}\label{sec-5}
In solid state physics, magnetic fields are experimentally controlled and modify band structures via the Landau quantization and the Zeeman splitting, which will be useful for tuning the qualitative property and quantitative strength of the Casimir effect.
For example, the zeroth Landau level (0LL) of Weyl fermions described by Eq.~(\ref{eq:WSMtoy}) is written as~\cite{Nguyen:2021}
\begin{align}
\omega^\mathrm{WSM-0LL} = -m + t^\prime (1-\cos ak_z) + \pi t^\prime \phi \label{eq:LLL}.
\end{align}
$\phi \equiv eBa_xa_y/h$ is the magnetic flux parallel to the $z$-direction with the magnetic-field strength $B$, the electric charge $e$, and the Planck constant $h$.
Note that, in Eq.~(\ref{eq:LLL}), we have neglected the Zeeman splitting, but it can be effectively included in the third term.
Within the parameter region of $-1<-m/ t^\prime + 1 + \pi\phi<1$, the 0LL in Eq.~(\ref{eq:LLL}) is distributed across Fermi level at the Fermi points.

Similarly, the 0LLs of Dirac electrons in Dirac semimetals can be described as~\cite{Nguyen:2021}
\begin{subequations}
\begin{align}
\omega^\mathrm{DSM-0LL}_\uparrow &= m - t^\prime (1-\cos ak_z) - \pi t^\prime \phi + \lambda_z g_\uparrow \phi \label{eq:spinup}, \\
\omega^\mathrm{DSM-0LL}_\downarrow &= -m + t^\prime (1-\cos ak_z) +\pi t^\prime \phi - \lambda_z g_\downarrow \phi \label{eq:spindown},
\end{align}
\label{eq:DSM_mag}
\end{subequations}
where the last terms mean the Zeeman splitting.

\begin{figure}[tb!]
    \centering
    \begin{minipage}[t]{0.33\columnwidth}
    \includegraphics[clip,width=1.0\columnwidth]{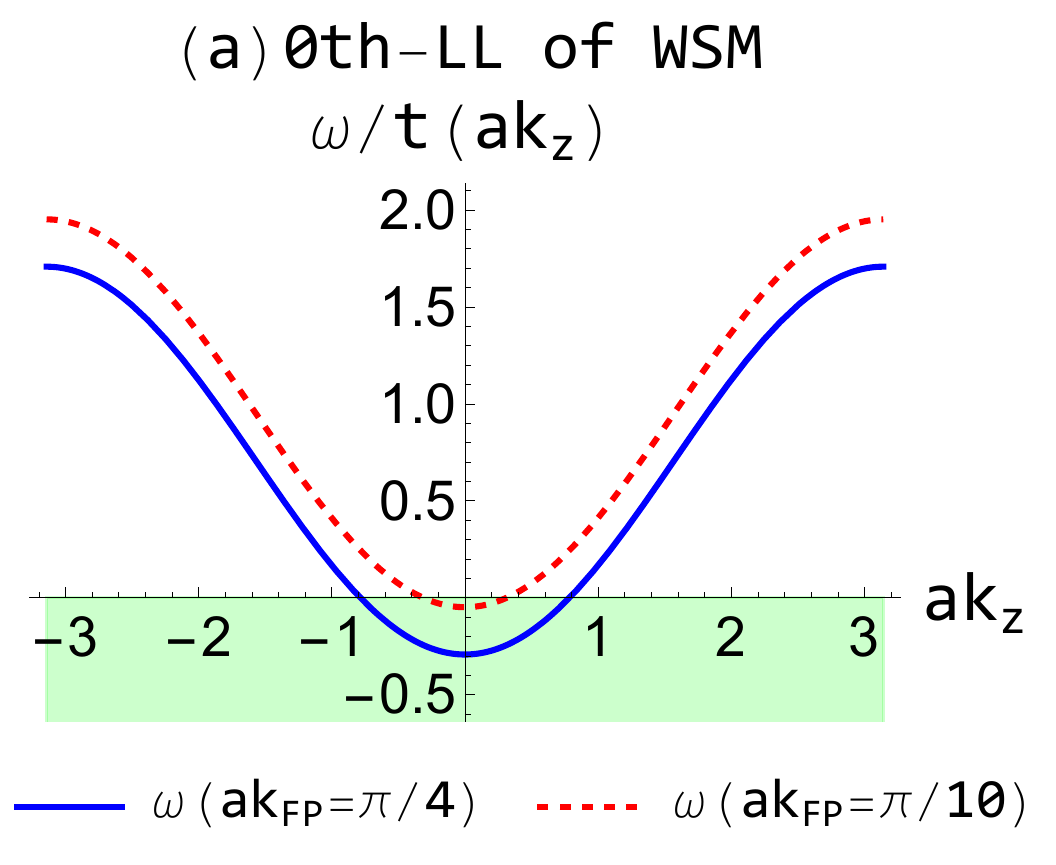}
    \end{minipage}%
    \begin{minipage}[t]{0.35\columnwidth}
    \includegraphics[clip,width=1.0\columnwidth]{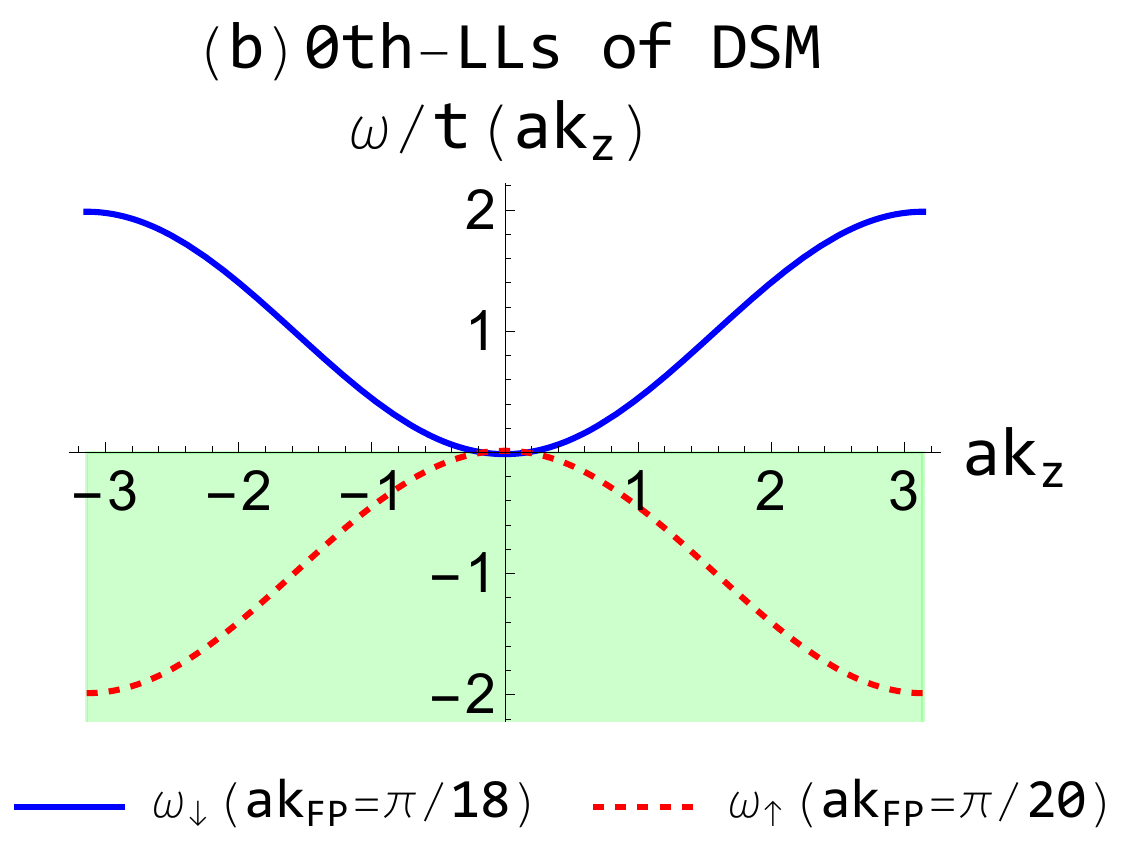}
    \end{minipage}
    \begin{minipage}[t]{0.5\columnwidth}
    \includegraphics[clip,width=1.0\columnwidth]{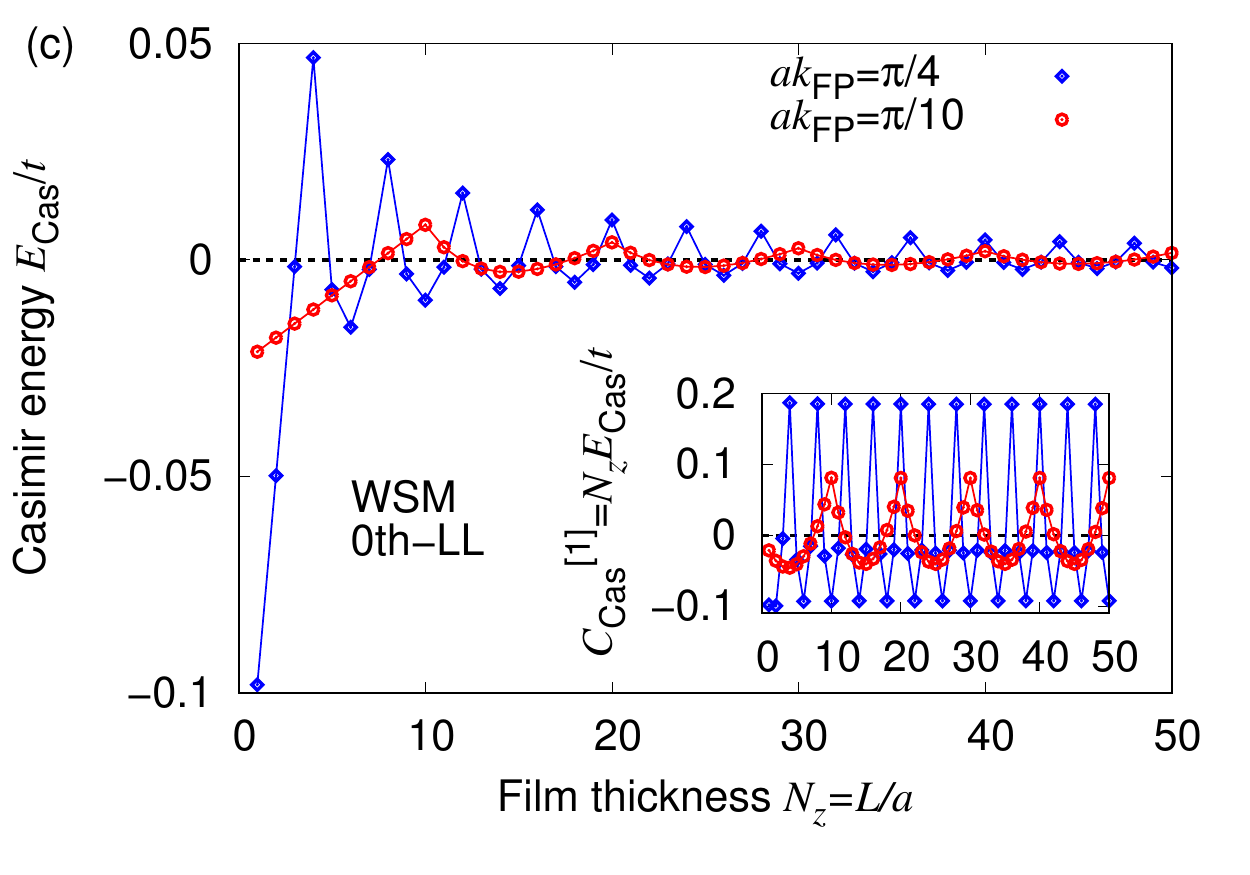}
    \end{minipage}%
    \begin{minipage}[t]{0.5\columnwidth}
    \includegraphics[clip,width=1.0\columnwidth]{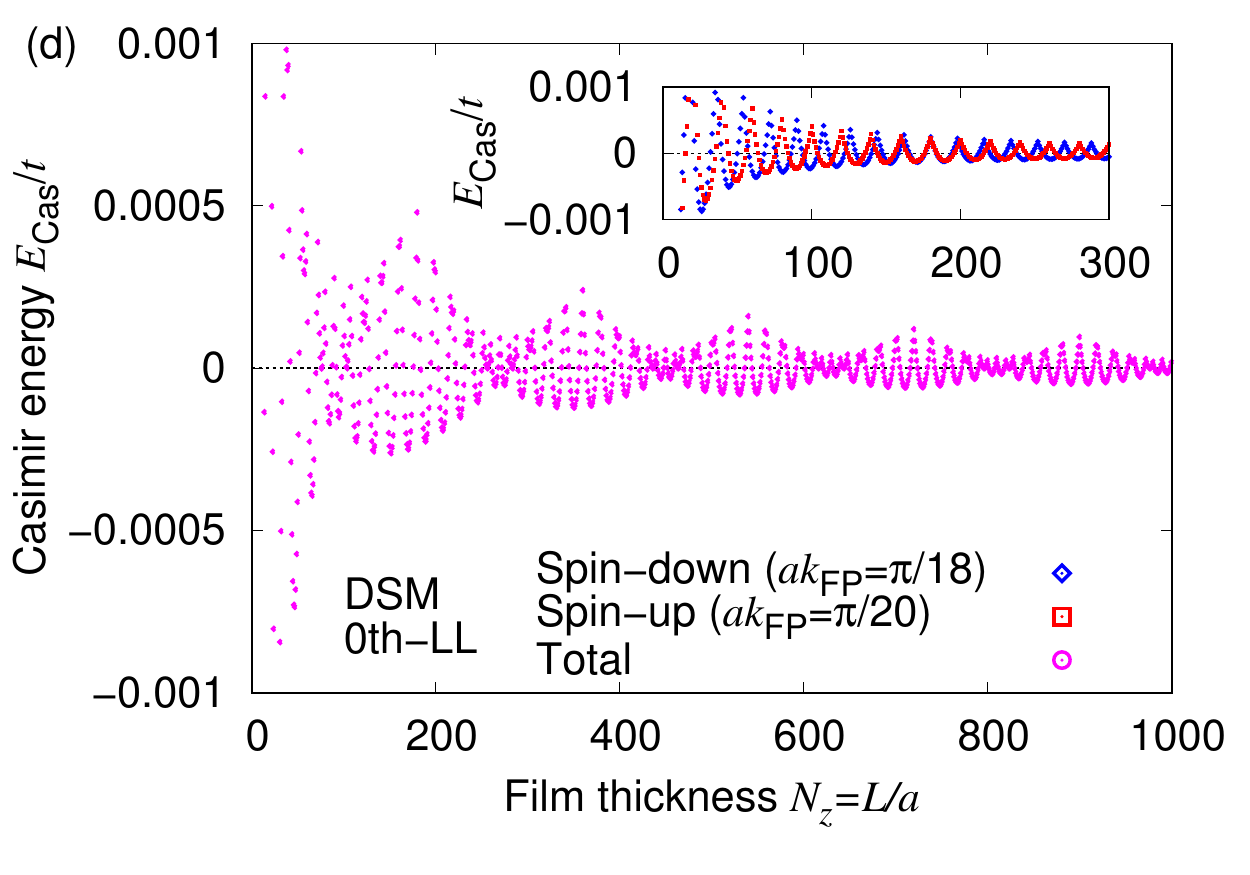}
    \end{minipage}
    \caption{(a)-(b) Dispersion relations of the zeroth Landau levels of Weyl/Dirac electrons in Weyl/Dirac semimetals, described by Eqs.~(\ref{eq:LLL}) and (\ref{eq:DSM_mag}).
(c)-(d) Casimir energy for Weyl/Dirac electrons with the phenomenological boundary (see Ref.~\cite{Nakayama:2022fvh} for details).
}
    \label{fig:mag}
\end{figure}

Figures~\ref{fig:mag}(c) shows the Casimir effect for the 0LL from Eq.~(\ref{eq:LLL}), where we fix the positions of the Fermi points (FPs) as $a k_\mathrm{FP} =\pi/4$ and $\pi/10$ [as shown in Fig.~\ref{fig:mag}(a)].
We find the oscillation of the Casimir energy, where the periods are $\tau_\mathrm{Cas} = 4$ and $10$, respectively.
Figure~\ref{fig:mag}(d) shows the results from Eqs.~(\ref{eq:DSM_mag}), where we fix the Fermi points of the spin-down and spin-up bands as $a k_\mathrm{FP} =\pi/18$ and $\pi/20$ [as shown in Fig.~\ref{fig:mag}(b)].
We find that the combination of the two periods induces a ``beat" of the Casimir energy, and its period is estimated as $\tau_\mathrm{beat} = 1/(\frac{1}{18} - \frac{1}{20}) = 180$.

\section{Nonrelativistic fields}\label{sec-6}
An open question regarding the Casimir effect is whether the {\it nonrelativistic} counterpart of the Casimir effect exists or not.
In the continuous spacetime, one may expect the absence of the Casimir energy by taking the heavy-mass limits of massive fields with the eigenvalues $\omega_\pm=\pm\sqrt{k^2+m^2}$ or by utilizing a renormalization scheme for quadratic dispersion relations with $\omega_\pm=\pm k^2/2m$.
Since nonrelativistic fields on the lattice are one of the most general band structures in solid state physics, the understanding of the Casimir effect for such fields will be essential.

Here, following Ref.~\cite{Nakayama:2022ild}, we demonstrate the Casimir effect for quadratic dispersion relations on the lattice, defined as
\begin{align}
&\omega_{\pm}^\mathrm{quad} = \pm \sum_{i}^{x,y,z} \frac{\hbar c}{a_i} (2-2\cos a_i k_i). \label{eq:quad}
\end{align}
For simplicity, we set $\hbar=1$, $c=1$, and $a \equiv a_x=a_y=a_z$.

\begin{figure}[tb!]
    \centering
    \begin{minipage}[t]{0.5\columnwidth}
    \includegraphics[clip,width=1.0\columnwidth]{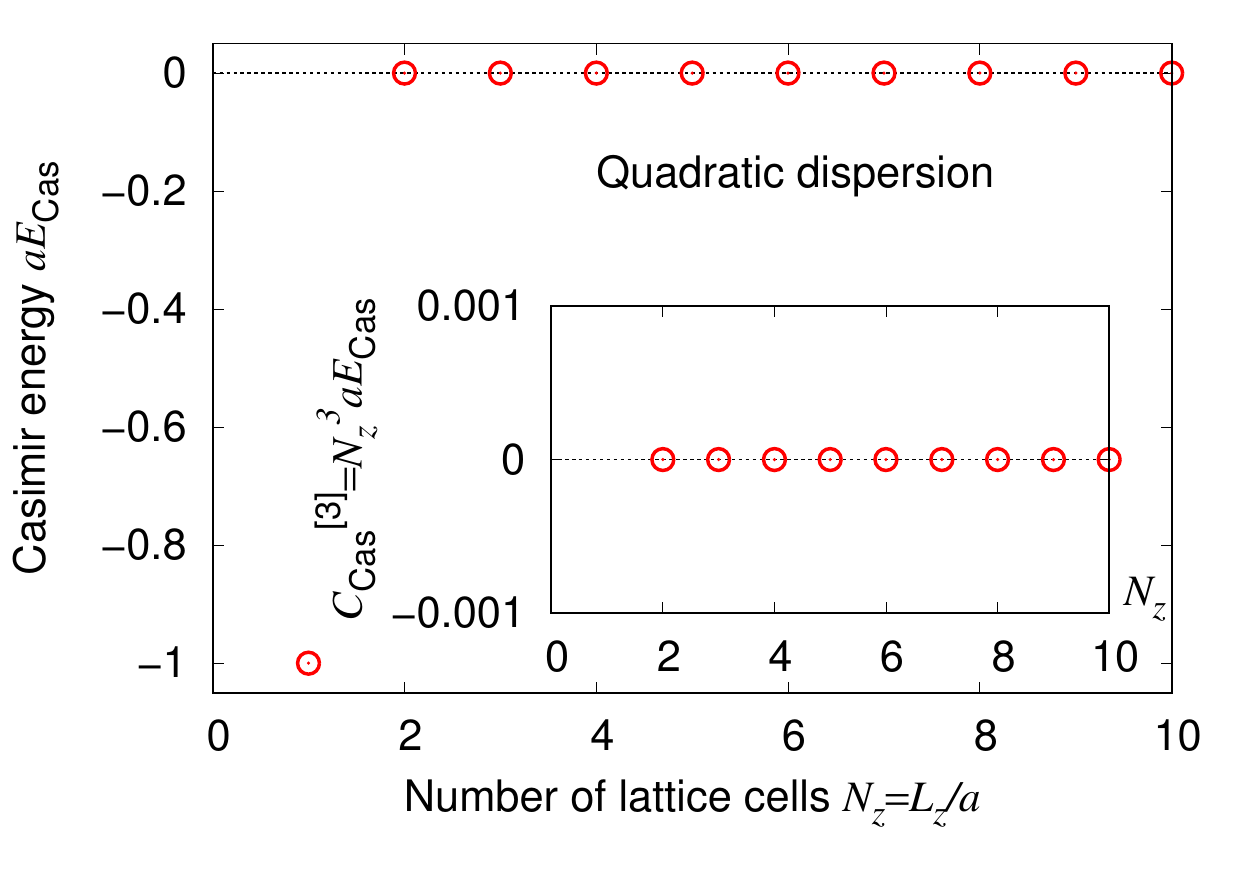}
    \end{minipage}
    \caption{Casimir energy for quadratic dispersion relations on the lattice (see Ref.~\cite{Nakayama:2022ild} for details).
}
    \label{fig:quad}
\end{figure}

Figure~\ref{fig:quad} shows the Casimir energy with the periodic boundary conditions.
From this figure, we find the Casimir energy survives only at $N_z=1$, while the Casimir energy is exactly zero at $N_z \geq 2$.
Such a disappearance of the Casimir effect at $N_z \geq 2$ is caused by an exact cancellation between $E_0^\mathrm{sum}$ and $E_0^\mathrm{int}$.
This behavior may be called the {\it remnant Casimir effect}~\cite{Nakayama:2022ild}.
Furthermore, for the phenomenological boundary~(\ref{eq:def_phenoBC}), we find that the Casimir energy is exactly zero at any $N_z$.

Note that such a behavior theoretically appears in various types of Hamiltonians.
For example, the 0LL of Weyl semimetals can be described as a cosine band as in Eq.~(\ref{eq:LLL}), and in the parameter region of $|-m/ t^\prime + 1 + \pi\phi| > 1$, this cosine band does not cross the Fermi level.
As a result, we can prove that the Casimir energy for such a band structure is exactly zero.

\section{Conclusions}\label{sec-8}
We have shown some examples of Casimir effects on the lattice.
In solid state physics, since there exist various types of relativistic or norelativistic quantum fields on the lattice, such as electrons, phonons, and magnons, the zero-point fluctuations from these fields induce the Casimir effect under a finite size of the system.
Such a study will open novel engineering fields utilizing the Casimir effect, which may be called Casimir electronics and Casimir spintronics.

\vspace{-10pt}
\section*{Acknowledgment}\label{sec-Ack}
\vspace{-10pt}
This work was supported by Japan Society for the Promotion of Science (JSPS) KAKENHI (Grants No. JP17K14277 and No. JP20K14476).

\vspace{-11pt}
\bibliographystyle{JHEP}
\bibliography{ref}
\end{document}